\documentclass[aps,prd,superscriptaddress,showkeys,showpacs,twocolumn,a4paper]{revtex4-1}
\usepackage{ulem}
\usepackage{color}
\usepackage{graphicx}
\usepackage{appendix}
\usepackage{epsfig}
\usepackage[colorlinks=true,linkcolor=blue]{hyperref}
\usepackage{epstopdf}
\usepackage{amsmath}
\usepackage{subfig}
\usepackage{comment}
\usepackage{float}
\usepackage[dvipsnames]{xcolor}
\newcommand{\be}{\begin{equation}}
\newcommand{\ee}{\end{equation}}
\newcommand{\ba}{\begin{eqnarray}}
\newcommand{\ea}{\end{eqnarray}}

\usepackage[cp1252]{inputenc}
\usepackage{eufrak}

\begin{document}
\title{Heavy quark transport coefficients in a viscous QCD medium with collisional and radiative processes}

\author{Adiba Shaikh}
\email{adibashaikh9@gmail.com}
\affiliation{Department of Physics, Indian Institute of Technology Bombay, Powai, Mumbai-400076, India}

\author{Manu Kurian}
\email{manu.kurian@iitgn.ac.in}
\affiliation{Indian Institute of Technology Gandhinagar, Gandhinagar-382355, Gujarat, India}

\author{Santosh K. Das}
%\email{santosh@iitgoa.ac.in}
\affiliation{School of Physical Sciences, Indian Institute of Technology Goa, Ponda-403401, Goa, India}

\author{Vinod Chandra}
%\email{vchandra@iitgn.ac.in}
\affiliation{Indian Institute of Technology Gandhinagar, Gandhinagar-382355, Gujarat, India}

\author{Sadhana Dash}
%\email{sadhana@phy.iitb.ac.in}
\affiliation{Department of Physics, Indian Institute of Technology Bombay, Powai, Mumbai-400076, India}

\author{Basanta K. Nandi}
%\email{basanta@phy.iitb.ac.in}
\affiliation{Department of Physics, Indian Institute of Technology Bombay, Powai, Mumbai-400076, India}

\begin{abstract}

The heavy quark drag and momentum diffusion coefficients in the presence of both the collisional and radiative processes have been studied in a hot viscous QCD medium. The thermal medium effects are incorporated by employing the effective fugacity quasiparticle model based on the lattice QCD equation of state. Viscous effects are embedded into the heavy quark transport through the near-equilibrium distribution functions of the constituent medium particles of the quark-gluon plasma. The viscous corrections to the momentum distributions have been estimated from the effective Boltzmann equation. The effect of shear viscous correction to drag and diffusion is investigated by considering the soft gluon radiation by heavy quarks along with the elastic collisional processes of the heavy quark with the light quarks and gluons within the QGP medium. The momentum and temperature dependence of the heavy quark transport coefficients are seen to be sensitive to the viscous coefficient of the QGP for the collisional and radiative processes. The collisional and radiative energy loss of the heavy quark in the viscous quark-gluon plasma has also been explored.

\end{abstract}

%\pacs{}

\keywords{Heavy quark, Drag and momentum diffusion, Energy loss, Gluon radiation, Viscous hydrodynamics}

\maketitle

%%%%%%%%%%%%%%%%%%%%%%%%%%%%%%%%%%%%%%%%%%%%%%%%%%%%%%
 \section{Introduction}
%%%%%%%%%%%%%%%%%%%%%%%%%%%%%%%%%%%%%%%%%%%%%%%%%%%%%%
Heavy-ion collision experiments at Relativistic Heavy Ion Collider (RHIC) at BNL and Large Hadron Collider (LHC) at CERN provided ample evidence of the formation of a new phase of hot and dense nuclear matter known as the quark-gluon plasma (QGP) with quarks, antiquarks and gluons as the fundamental degrees of freedom~\cite{Adams:2005dq, Back:2004je,Arsene:2004fa,Adcox:2004mh,Aamodt:2010pb,Jaiswal:2020hvk}. The QGP evolution has been successfully described within the framework of relativistic viscous hydrodynamics~\cite{Heinz:2013th,Gale:2013da, Jaiswal:2016hex}. The dissipative processes in the QGP and the associated transport coefficients are sensitive to the medium evolution. Various transport coefficients associated with the transport processes in the hot QCD/QGP medium can be determined from the underlying microscopic theories (QCD or effective kinetic theory approach). They could also be extracted from the experimental observables at the RHIC and LHC. Previous studies with viscous hydrodynamics focused on a small value of shear viscosity to entropy density ratio ($\eta/\mathfrak{s}$)~\cite{Romatschke:2007mq}, and recently the impact of bulk viscosity on the evolution of the QGP have been explored~\cite{Ryu:2015vwa}.

Heavy quarks, mainly charm and bottom, are created dominantly due to partonic hard scattering in the early stages of the heavy-ion collisions and are considered as an effective probe to study the QGP properties~\cite{Rapp:2018qla,Prino:2016cni,Dong:2019unq,Andronic:2015wma,Aarts:2016hap}. Due to their large masses ($m_c \approx  1.3$ GeV, $m_b \approx  4.2$ GeV) in comparison to the temperature of the thermal background medium ($T$), they traverse through the QGP without being equilibrated with the medium constituents and thus carry information about the evolution of the QGP. The heavy quark dissipates energy while traveling through the QGP via collisional process (elastic interaction) and through the radiative process (inelastic interaction)~\cite{Braaten:1991we,Mustafa:2004dr,Mazumder:2011nj,Abir:2012pu,Cao:2013ita,Sarkar:2018erq,Cao:2016gvr,Zigic:2018ovr,Liu:2020dlt,Uphoff:2014hza,Gossiaux:2006yu,Das:2010tj,Cao:2015hia}. The Brownian motion of heavy quarks in the QGP medium can be described within the framework of the Fokker-Planck dynamics, where the interactions of the heavy quarks with the medium constituents are incorporated through the drag and momentum diffusion coefficients~\cite{Svetitsky:1987gq,GolamMustafa:1997id}. Several studies have been performed to explore the heavy quark transport coefficients, and the associated measured observables in heavy-ion collisions such as nuclear modification factor $R_{AA}$, directed and elliptic flow coefficients in the hot QCD medium~\cite{Cao:2018ews,Das:2013kea,vanHees:2005wb,Gossiaux:2008jv,Xu:2018gux,Das:2015ana,Scardina:2017ipo,Song:2015sfa,Adler:2005xv,Adare:2006nq,Alberico:2013bza,vanHees:2007me,Li:2019wri,Jamal:2021btg,Akamatsu:2008ge,Uphoff:2011ad,Xu:2013uza,Banerjee:2011ra,Brambilla:2020siz}. 
%However, many investigations considered the elastic interaction of the heavy quark in the medium. 
The impact of soft gluon radiation by the heavy quark on its transport coefficients in the thermalized QGP medium has been recently explored in Refs.~\cite{Mazumder:2013oaa,Prakash:2021lwt}. It is observed that the collisional energy loss is dominant at the low momentum regime of the heavy quark, whereas at high momentum regimes, energy loss due to medium induced gluon radiation by the heavy quark is dominant. The heavy quark transport coefficients, while considering the radiative process of heavy quarks along with the collisional process in a viscous QGP medium, is an interesting aspect to explore, and this sets the motivation for the present study.

The current focus is to study the sensitivity of the heavy quark drag and diffusion coefficients to the shear viscosity for the collisional and radiative energy loss in viscous QGP. The realistic equation of state effects is embedded in the analysis through the effective fugacity quasiparticle model (EQPM)~\cite{Chandra:2011en,Chandra:2008kz} description of the QGP medium. The non-equilibrium distribution function has been obtained by solving the consistently developed effective Boltzmann equation based on the EQPM by employing the Chapman-Enskog like iterative method within the relaxation time approximation (RTA) ~\cite{Mitra:2018akk}. Notably, the mean field contributions that originate from the basic conservation laws are incorporated in the estimation of the near-equilibrium momentum distribution functions. The impact of viscous coefficients of the QGP has already been explored in photon production, dilepton emission, and many relevant observables of heavy-ion collisions~\cite{Shen:2014nfa,Dusling:2008xj,Vujanovic:2013jpa,Thakur:2020ifi,Chandra:2015rdz}. Recently performed estimations of heavy quark dynamics in the anisotropic medium have investigated the effects of momentum anisotropy of the QGP on its transport coefficients, energy loss, and the associated nuclear modification factor $R_{AA}$~\cite{Chandra:2015gma,Prakash:2021lwt}. In Refs.~\cite{Das:2017dsh,Mrowczynski:2017kso,Ruggieri:2018rzi,Boguslavski:2020tqz,Carrington:2020sww,Sun:2019fud}, the physics of non-equilibrium dynamics of the medium to the heavy quark transport and related observables have been investigated. The viscous corrections to the heavy quark transport coefficients due to the collisional processes have been studied in Refs.~\cite{Das:2012ck,Kurian:2020orp,Song:2019cqz,Singh:2019cwi,Kurian:2020kct}. 

The prime focus of this work has been to investigate the impact of the shear viscosity of the QGP on the radiative processes regarding heavy quark dynamics. In this context, heavy quark drag and diffusion coefficients along with the heavy quark energy loss have been studied and analyzed in contrast to that from the collisional ($2\rightarrow 2$ scattering) process. The shear viscous corrections have been observed to have a sizable impact on the heavy quark transport coefficients. 

The article is organized as follows. In section II, the formulation of the heavy quark dynamics in the viscous QGP medium is discussed while incorporating the collisional and radiative processes along with the EQPM description of viscous corrections to the quarks, antiquarks, and gluon distribution functions. Section III is devoted to the results and discussions. The analysis is summarized with an outlook in section IV.\\

\noindent {\bf Notations and conventions}: In the article, the subscript $k$ denotes the particle species, $i.e.$, $k=(lq, l\bar{q},g)$, with $lq$, $l\bar{q}$ and $g$ representing light quarks, light antiquarks and gluons, respectively. The degeneracy factor for gluon is $\gamma_g=N_s\times (N_c^2-1)$ and for light quark (antiquark) is $\gamma_{lq}=N_s\times N_c\times N_f$ with $N_s=2$, $N_f=3$ ($u,d,s$), and $N_c=3$ (for $SU(3)$). The quantity $u^{\mu}$ is the normalized fluid velocity with  $u^{\mu}u_{\mu}=1$ and $g^{\mu\nu}=\text{diag}(1, -1, -1, -1)$ is the metric tensor.

%%%%%%%%%%%%%%%%%%%%%%%%%%%%%%%%%%%%%%%%%%%%%%
\section{Formalism}
%%%%%%%%%%%%%%%%%%%%%%%%%%%%%%%%%%%%%%%%%%%%%%

%
%%%%%%%%%%%%%%%%%%%%%%%%%%%%%%%%%%%%%%%%%%%%%%%%%%%%%%%
\subsection{Heavy quark transport coefficients}
%%%%%%%%%%%%%%%%%%%%%%%%%%%%%%%%%%%%%%%%%%%%%%%%%%%%%%% 
%
Heavy quarks can be considered to be a non-equilibrated degree of freedom executing Brownian motion within the background QGP medium.  Such massive quarks lose their energy due to collision with the medium constituents (elastic 2$\rightarrow$2 process) and through gluon radiation (inelastic 2$\rightarrow$3 process). Both these interactions of heavy quarks in the QGP medium are embedded in the drag and diffusion coefficients. We initiate the analysis with the collisional process followed by the inelastic radiative process.

%
%%%%%%%%%%%%%%%%%%%%%%%%%%%%%%%%%%%%%%%%%%%%%%%%%%%%%%%%
\subsubsection{Collisional process}
%%%%%%%%%%%%%%%%%%%%%%%%%%%%%%%%%%%%%%%%%%%%%%%%%%%%%%%%  
%

While traversing through the QGP medium, heavy quark $(HQ)$ undergoes collisions with the medium constituents, $i.e.$, light quarks $(lq)$, light antiquarks $({\bar lq})$ and gluons {$(g)$}. Here, we consider the elastic ($2\rightarrow 2$) process,
\begin{equation}
    HQ \ (p)+lq/l\bar{q}/g \ (q) \rightarrow HQ \ (p')+lq/l\bar{q}/g  \ (q').\label{Col}
\end{equation}  
We followed the formalism developed in~\cite{Svetitsky:1987gq} to study the Brownian motion of heavy quarks in the medium. The Boltzmann transport equation for the evolution of the heavy quark momentum distribution $f_{HQ}$ reduces to the Fokker-Planck equation within the soft scattering approximation and has the following form, 
\begin{align}\label{1.1}
  	\frac{\partial f_{HQ}}{\partial t}=\frac{\partial}{\partial p_i}\left[A_i({\bf p})f_{HQ}+\frac{\partial}{\partial p_j}\Big(B_{i j}({\bf p})f_{HQ}\Big)\right],
\end{align}
where the drag force $A_i({\bf p})$ and momentum diffusion $B_{ij}({\bf p})$ of the heavy quark respectively take the forms,
\begin{align}
   A_i({\bf p}) =& \,\frac{1}{2 E_p  \gamma_{HQ}} \int \frac{d^3 {\bf q}}{(2 \pi)^3 E_q}  \int \frac{d^3 {\bf q'}}{(2 \pi)^3 E_{q'}} \int \frac{d^3 {\bf p'}}{(2 \pi)^3 E_{p'}}\nonumber\\
   &\times\sum{|{\mathcal{M}}_{2\rightarrow2}|^2} \ (2 \pi)^4 \  \delta^{(4)}(p+q-p'-q') \nonumber\\
   & \times f_k(E_q) \  (1\pm f_k(E_{q'})) \ [(p-p')_i]\nonumber\\
   =& \ \langle\langle ( p - p')_i \rangle\rangle, \label{Ai}
\end{align}
and
\begin{align}
   B_{ij}({\bf p}) =& \frac{1}{2 E_p  \gamma_{HQ}} \int \frac{d^3 {\bf q}}{(2 \pi)^3 E_q}  \int \frac{d^3 {\bf q'}}{(2 \pi)^3 E_{q'}} \int \frac{d^3 {\bf p'}}{(2 \pi)^3 E_{p'}}\nonumber\\
   &\times\sum{|{\mathcal{M}}_{2\rightarrow2}|^2} \ (2 \pi)^4 \  \delta^{(4)}(p+q-p'-q') \nonumber\\
   & \times f_k(E_q) \ (1\pm f_k(E_{q'})) \ [(p-p')_i]\nonumber\\
   =& \frac{1}{2} \langle\langle (p-p')_i \ ( p - p')_j \rangle\rangle, \label{Bij}
\end{align}
where $\gamma_{HQ}=N_s \times N_c$ is the heavy quark degeneracy factor. The term $|{\mathcal{M}}_{2\rightarrow2}|$ represents the scattering amplitude of the collisional process for $2\rightarrow2$ process as depicted in Fig.~\ref{Feyn_col} and is described in Appendix~\ref{A1}. Here, $f_k(E_q)$ denotes the Fermi-Dirac distribution function for quarks and the Bose-Einstein distribution function for gluons. Incorporating quantum statistics, we have considered Fermi suppression $(1 - f_{lq}(E_{q'}))$ and Bose enhancement $(1 + f_{g}(E_{q'}))$ for the final state phase space of the quarks and gluons, respectively. The position dependence of $f_{HQ}({\bf p},t)$ is neglected by assuming its homogeneity with respect to spatial coordinate. The drag force measures the thermal average of the momentum transfer, whereas $B_{ij}$ quantifies the square of the momentum transfer due to the interactions of heavy quarks in the medium. As both $A_i({\bf p})$ and $B_{ij}({\bf p})$ depend only on the initial heavy quark momentum $({\bf p})$, the drag force and momentum diffusion of the heavy quarks can be decomposed as follows,
\begin{align}
    A_i =& \ p_i \ A(p^2),\label{Aip}\\
    B_{ij} =& \left[ \delta_{ij} - \frac{p_ip_j}{p^2} \right]B_0(p^2) + \frac{p_ip_j}{p^2} B_1(p^2),\label{Bijp} 
\end{align}
where $p=|{\bf p}|$ is the magnitude of heavy quark initial momentum. From Eq.(\ref{Aip}), the heavy quark drag coefficient is defined as,
\begin{equation}
A=\langle\langle 1 \rangle\rangle - \frac{\langle\langle {\bf{p.p'} \rangle\rangle}}{p^2}.\label{Ap}
\end{equation}
Similarly, the transverse and longitudinal momentum diffusion coefficients are defined as, 
\begin{align}
&B_{0}= \frac{1}{4}\left[\langle\langle p'^{2} \rangle\rangle-\frac{\langle\langle ({\bf{p.p'}})^2\rangle\rangle}{p^2} \right],\label{B0p}\\ 
&B_{1}= \frac{1}{2}\left[\frac{\langle\langle ({\bf{p.p'})}^2\rangle\rangle}{p^2} -2\langle\langle ({\bf{p.p'})}\rangle\rangle +p^2 \langle\langle 1 \rangle\rangle\right]\label{B1p},
\end{align}
respectively. The kinematics is simplified in the center-of-momentum frame of the system, and the thermal average of a function $F({\bf p})$ for $2\rightarrow 2$ process in the center-of-momentum frame takes the form as follows,
\begin{align}
    \langle \langle F({\bf p})\rangle \rangle_{col}&=\frac{1}{(512 \, \pi^4) E_p \gamma_{HQ}}\int_0^\infty dq \left(\frac{s-m_{HQ}^2}{s}\right) f_k(E_q)\nonumber \\ &\times (1\pm f_k(E_{q'})) \int_0^\pi d\chi \, \sin\chi \int_0^\pi d\theta_{cm} \, \sin\theta_{cm} \nonumber\\ &\times\sum{|{\mathcal{M}}_{2\rightarrow2}|^2} \int_0^{2\pi} d\phi_{cm} \ F({\bf p}),\label{Fcm}
\end{align}
where $\chi$ is the angle between the incident heavy quark and medium particles in the lab frame. Here, $\theta_{cm}$ and $\phi_{cm}$ is the zenith and azimuthal angle in the center-of-momentum frame, respectively. The Mandelstam variables ($s,t,u$) are defined as,
\begin{align}
s = & (E_p+E_q)^2-(|{\bf p}|^2+|{\bf q}|^2+2|{\bf p}||{\bf q}| \sin \chi), \\
t = & 2 \, p_{cm}^2 (\cos \theta_{cm}-1),\\
u = & 2 \, m_{HQ}^2-s-t,
\end{align}
where $p_{cm}=|{\bf p}_{cm}|$ represents the magnitude of heavy quark initial momentum in center-of-momentum frame. It is important to note that the IR divergences occurring due to $t$-channel gluonic propagator in Fig.~\ref{Feyn_col} $(a)$ and $(d)$ is regularized by inserting the Debye screening mass ($m_D$) at leading order for gluons within the medium.

\begin{figure}[h]
\includegraphics[scale=1]{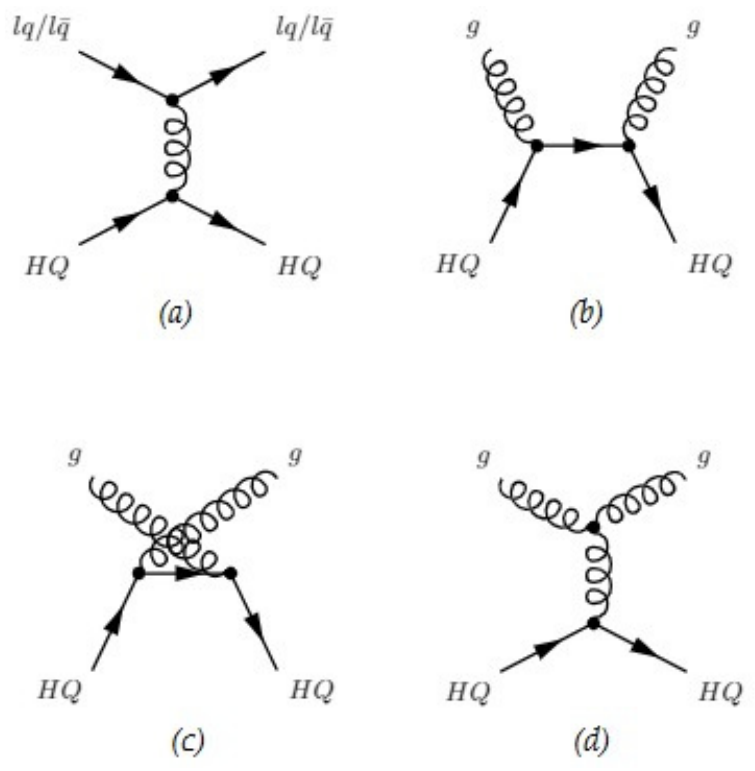} 
\caption{\small Heavy quark $2 \rightarrow 2$ processes with ($a$) $lq/l{\bar q}$ (t-channel), ($b$) $g$ ($s$-channel), ($c$) $g$ ($u$-channel), ($d$) $g$ ($t$-channel).}
\label{Feyn_col}
\end{figure}
%%%%%%%%%%%%%%%%%%%%%%%%%%%%%%%%%%%%%%%%%%%%%%%%
\subsubsection{Radiative process}
\begin{figure}[h]
\includegraphics[scale=1]{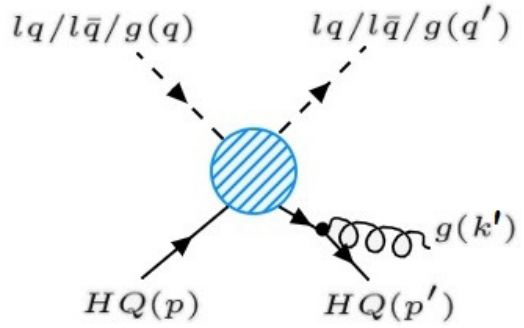} 
\caption{\small Partonic $2 \rightarrow 3$ process considered for \\ inelastic collision of $HQ$ with $lq/l{\bar q}/g$ and a soft gluon emission in the final state. Here, the blob represents all the $2 \rightarrow 2$ processes displayed in Fig.~\ref{Feyn_col}.}
\label{feyn_rad}
\end{figure}

%%%%%%%%%%%%%%%%%%%%%%%%%%%%%%%%%%%%%%%%%%%%%%%%
%
Heavy quarks can radiate gluons while moving through the QGP medium along with collisions with the medium constituents. We consider the inelastic ($2\rightarrow 3$) process,
\begin{equation}
    HQ \ (p)+lq/l\bar{q}/g \ (q) \rightarrow HQ \ (p')+lq/l\bar{q}/g  \ (q') + g \ (k'),\label{Rad}
\end{equation}  
where $k'\equiv(E_{k'},{\bf k'_{\perp}},k'_z)$ is the four-momentum of the emitted soft gluon by the heavy quark in the final state ($k'\rightarrow 0$). Being an inelastic process, only the kinematical and the interaction parts change in comparison to Eqs. (\ref{Ai}) and (\ref{Bij}) and the transport part remains the same.
The general expression for the thermal averaged $F({\bf p})$ for $2\rightarrow 3$ process is as follows~\cite{Mazumder:2013oaa},
\begin{align}
    \langle \langle F({\bf p}) &\rangle \rangle_{rad} =\frac{1}{2 E_p  \gamma_{HQ}} \int \frac{d^3 {\bf q}}{(2 \pi)^3 E_q}  \int \frac{d^3 {\bf q'}}{(2 \pi)^3 E_{q'}} \int \frac{d^3 {\bf p'}}{(2 \pi)^3 E_{p'}} \nonumber\\ &\times\int \frac{d^3 {\bf k'}}{(2 \pi)^3 E_{k'}} \sum{|{\mathcal{M}}_{2\rightarrow 3}|^2} \  \delta^{(4)}(p+q-p'-q'-k') \nonumber\\&\times (2 \pi)^4 \ f_k(E_q) \ (1 \pm f_k(E_{q'})) \ (1 + f_k(E_{k'})) \ \nonumber\\& \times \theta_1(E_p-E_{k'}) \ \theta_2(\tau-\tau_F) \ F({\bf p}).\label{Frad}
\end{align}
The theta function $\theta_1(E_p-E_{k'})$ constraints the phase space where heavy quark initial state energy $E_p$ is always greater than the radiated soft gluon energy $E_{k'}$ in the final state and $\theta_2(\tau-\tau_F)$ ensures that the collision time $\tau$ of the heavy quark with the medium particles is greater than the gluon formation time $\tau_F$ (Landau-Pomeranchuk-Migdal Effect)~\cite{Wang:1994fx,Gyulassy:1993hr,Klein:1998du}. The Bose enhancement factor $(1 + f_g(E_{k'}))$ is for the radiated gluon in the final state. The term $|{\mathcal{M}}_{2\rightarrow 3}|^2$ denotes the matrix element squared for $2 \rightarrow 3$ radiative process as depicted in Fig.~\ref{feyn_rad}, which can be expressed in terms of the collision process multiplied by the probability for soft gluon emission \cite{Abir:2011jb} as,
\begin{equation}\label{M23}
    |{\mathcal{M}}_{2\rightarrow 3}|^2=|{\mathcal{M}}_{2\rightarrow 2}|^2 \times  \frac{12 g_s^2}{k'_\perp} \left(1+\frac{m_{HQ}^2}{s}e^{2y_{k'}}\right)^{-2}, 
\end{equation}
where $y_{k'}$ is the rapidity of the emitted gluon and 
$\left(1+\frac{m_{HQ}^2}{s}e^{2y_{k'}}\right)^{-2}$ is the dead cone factor for the heavy quark. In the limit of soft gluon emission ($\theta_{k'}<<1$) we have,
\begin{equation}\label{dcf}
    \left(1+\frac{m_{HQ}^2}{s}e^{2y_{k'}}\right)^{-2} \approx \left(1+\frac{4 \, m_{HQ}^2}{s \, \theta_{k'}^2 }\right)^{-2},
\end{equation}
where $\theta_{k'}$ is the angle between the heavy quark and the radiated soft gluon which is related to its rapidity by $y_{k'}=-\,\ln[\tan(\theta_{k'}/2)]$. From Eqs.~(\ref{M23}) and~(\ref{dcf}), the heirarchy in the radiative energy loss for light quarks ($lq$), charm ($c$) and bottom ($b$) is,
\begin{center}
 $|{\mathcal{M}}_{2\rightarrow 3}|_{b}<|{\mathcal{M}}_{2\rightarrow 3}|_{c}<|{\mathcal{M}}_{2\rightarrow 3}|_{lq},$
\end{center}
where $m_{b}>m_{c}>m_{lq}$. The heavy quark transport coefficient for the radiative process in Eq.~(\ref{Frad}) can be further simplified in terms of the kinematic part of Eq.~(\ref{Fcm}) in the center-of-momentum frame as follows,
\begin{widetext}
\begin{align} \label{rad}
    \langle \langle F({\bf p}) \rangle \rangle_{rad} =&\frac{1}{(512 \, \pi^4)E_p \gamma_{HQ}} \int_0^\infty dq \left(\frac{s-m_{HQ}^2}{s}\right) f_k(E_q) \  (1\pm f_k(E_{q'})) \int_0^\pi d\chi \ \sin\chi\int_0^\pi d\theta_{cm} \ \sin{\theta_{cm}}\int_0^{2\pi} d\phi_{cm}  \nonumber \\ &   \times  \int \frac{d^3 k'}{(2 \pi)^3 2 E_{k'}}  \frac{12  g_s^2}{k'_\perp} \  \left(1+\frac{m_{HQ}^2}{s}e^{2y_{k'}}\right)^{-2} (1 + f_g(E_{k'})) \ \theta_1(E_p-E_{k'}) \ \theta_2(\tau-\tau_F) \  \sum{|{\mathcal{M}}_{2\rightarrow2}|^2} \ F({\bf p}).
\end{align}
\end{widetext}
The evaluation of the soft gluon 3-momentum integral is discussed in detail in Appendix~\ref{A2}.

%%%%%%%%%%%%%%%%%%%%%%%%%%%%%%%%%%%%%%%%%%%%%
\subsection{EQPM distribution of quarks and gluons in a viscous medium}
%%%%%%%%%%%%%%%%%%%%%%%%%%%%%%%%%%%%%%%%%%%%%
%
An adequate modelling of the viscous QGP medium is needed for the effective description of heavy quark transport while including the effects of the thermal interactions of the medium via realistic QCD equation of state. To that end, we employ the EQPM in the analysis. For the near-equilibrium system (not very far from local equilibrium), the particle momentum distribution function takes the following form, 
\begin{align}\label{2.1}
 &f_k=f^0_k+\delta f_k,   &&\delta f_k/f^0_k \ll1,
\end{align} 
with $f^0_k$ as the EQPM equilibrium distribution function. 
The EQPM distribution functions of light quarks/antiquarks and gluons at vanishing baryon chemical potential can be defined in terms of effective fugacity parameter $z_k$ to encode the QCD medium interactions as follows,
\begin{align}\label{2.2}
&f^0_{lq/l\bar{q}} =\frac{z_{lq} \exp[-\beta (u\!\cdot\! q)]}{1 + z_{lq}\exp[-\beta (u\!\cdot\! q)]},\\
&f^0_g =\frac{z_g \exp[-\beta\, (u\!\cdot\! q)]}{1 - z_g\exp[-\beta\, (u\!\cdot\! q)]}.\label{2.202}
\end{align}
The realistic hot QCD medium equation of state can be interpreted in terms of non-interacting quasiparticles having temperature-dependent effective fugacities.
The EQPM description of the QCD medium was seen to be thermodynamically consistent, realizing the medium as a grand canonical ensemble of quarks/antiquarks and gluons. The effective grand canonical partition function for the QGP medium $Z_{{eff}}$, which yields the above forms of the equilibrium EQPM distribution is as follows~\cite{Chandra:2011en,Chandra:2008kz},
\begin{equation}
    Z_{{eff}}=Z_gZ_{lq}Z_{l\bar{q}},
\end{equation}
where 
\begin{equation}
    \ln Z_{k}=\pm \gamma_kV\int \frac{d\mid{\bf{{q}}}_k\mid}{(2\pi)^3}\ln (1\pm z_k\exp{(-\beta (E_q-a_k \mu)}),
\end{equation}
are the quark/antiquark and gluonic contributions, respectively. Here,  $V$ is the volume, $-$ is for the gluonic and $+$ is for the quark case. The temperature behaviour of the fugacity parameter can be obtained by fitting the pressure obtained within the EQPM description ($P\beta V=\ln Z_{{eff}}$) with the lattice QCD results, see Ref.\cite{Chandra:2008kz} for more details.

The physical significance of $z_k$ can be understood from  the single-particle energy dispersion. From the fundamental thermodynamic relation, we can define the quasiparticle energy $\omega_k$ as follows,
\begin{equation}
\omega_k=-\frac{1}{V}\partial_\beta\ln Z_{{eff}}=E_q+\delta\omega_k,
\end{equation}
with $\delta\omega_k=T^2\partial_T\ln({z_k})$ as the medium modified part of the dispersion relation. Hence, the fugacity parameter modifies the covariant form of the dispersion relation as follows~\cite{Mitra:2018akk},
\begin{equation}\label{2.3}
\Tilde{q_k}^{\mu} = q_k^{\mu}+\delta\omega_k\, u^{\mu}, 
\end{equation}
where $\Tilde{q}_k^{\mu}=(\omega_k, {\bf q}_k)$ and $q_k^{\mu}=(E_q, {\bf q}_k)$ are the dressed (quasiparticle) and bare particle momenta, respectively. Note that at the limit $z_k\rightarrow 1$, the system approaches the ultra-relativistic limit (ideal equation of state) and the medium modified part of the energy dispersion vanishes, $i.e.$, $\delta\omega_k\rightarrow 0$.
Further, one can define an effective coupling from the kinetic theory following the definition in terms of EQPM momentum distributions. The EQPM description of the hot QCD medium is based on the charge
renormalization in medium~\cite{Chandra:2011en}, and can be realized in terms of the effective coupling $\alpha_{eff}$ as~\cite{Mitra:2017sjo},
\begin{equation}\label{2.4}
\frac{\alpha_{eff}}{\alpha_s(T)}= \dfrac{\frac{2N_c}{\pi^2}\mathrm{PolyLog}~[3,z_g]
-\frac{2N_f}{\pi^2}\mathrm{PolyLog}~[3,-z_{lq}]}{\left(\frac{N_c}{3}+\frac{N_f}{6}\right)}.
\end{equation}
Here, $\alpha_{s}(T)$ denotes the 2-loop running coupling constant at finite temperature and has the form~\cite{Kaczmarek:2005ui,Das:2015ana},
\begin{equation}\label{alphas}
    \alpha_{s}(T)=\frac{1/4\pi}{\beta_0 \left[\log\left(\frac{2\pi T}{1.3T_c}\right)^2\right]+\left(\frac{\beta_1}{\beta_0}\right) \log\left[\log\left(\frac{2\pi T}{1.3T_c}\right)^2\right]},
\end{equation}
where $\beta_0$ and $\beta_1$ respectively take the forms,
\begin{align}\label{2.5}
    &\beta_0=\frac{11N_c-2N_f}{48\pi^2},
     &&\beta_1=\frac{102N_c-38N_f}{3(16\pi^2)^2}.
\end{align}
The utility of the model in the context of the QGP (hot QCD medium) has been realized by setting up an effective kinetic theory. The evolution of the distribution function is described by the effective Boltzmann equation based on the EQPM. The covariant form of the effective transport equation within the RTA is as follows~\cite{Mitra:2018akk},
\begin{equation}\label{2.6}
\Tilde{q}^{\mu}_k\,\partial_{\mu}f_k(x,\Tilde{q}_k)+F_k^{\mu}\left(u\!\cdot\!\tilde{q}_k\right)\partial^{(q)}_{\mu} f_k = -\left(u\!\cdot\!\tilde{q}_k\right)\frac{\delta f_k}{\tau_R},
\end{equation}
where $\tau_{R}$ is the thermal relaxation time and  $F_k^{\mu}=-\partial_{\nu}(\delta\omega_k u^{\nu}u^{\mu})$ denotes the mean field force term that originates from the conservation laws of energy-momentum and particle flow in the medium. The viscous corrections to the distribution function are obtained by solving Eq.~(\ref{2.6}). We adopt an iterative Chapman-Enskog like method~\cite{Jaiswal:2013npa} for solving the relativistic Boltzmann equation and obtain,
\begin{align}\label{2.7}
\!\!\delta f_k = \tau_R\bigg( \Tilde{q}_k^\gamma\partial_\gamma \beta + \frac{\beta\, \Tilde{q}_k^\gamma\, \Tilde{q}_k^\phi}{u\!\cdot\!\Tilde{q}_k}\partial_\gamma u_\phi -\beta\theta\,\delta\omega_k \bigg) f^0_k\Tilde{f}^0_k, 
\end{align}
where $\tilde{f_k}^0\equiv(1-a_kf_k^0)$ ($a_{g}=-1$ for bosons and $a_{lq}=+1$ for fermions). The first-order evolution for the shear stress tensor $\pi^{\mu\nu}$ within the effective kinetic theory has the following forms~\cite{Bhadury:2019xdf},
\begin{align}\label{2.8}
\pi^{\mu\nu} = 2\,\tau_R\,\beta_\pi\,\sigma^{\mu\nu},
\end{align}
with $\theta\equiv\partial_{\mu}u^{\mu}$ as the scalar expansion and $\sigma^{\mu\nu}\equiv\Delta^{\mu\nu}_{\alpha\beta}\nabla^{\alpha}u^{\beta}$ where $\Delta^{\mu\nu}_{\alpha\beta}\equiv\frac{1}{2}(\Delta^\mu_\alpha\Delta^\nu_\beta +\Delta^\mu_\beta\Delta^\nu_\alpha)-\frac{1}{3}\Delta^{\mu\nu}\Delta_{\alpha\beta}$ denotes as the traceless symmetric projection operator orthogonal to the fluid velocity $u^{\mu}$.
Here, $\beta_\pi$ is the first-order coefficient and has the form,
\begin{align}\label{2.9}
\beta_\pi=~ \beta \sum_k\bigg[\Tilde{J}^{(1)}_{k~42}+(\delta\omega_k)\Tilde{L}^{(1)}_{k~42}\bigg].
\end{align}
The thermodynamic integrals $\Tilde{J}^{(r)}_{k~nm}$ and $\Tilde{L}^{(r)}_{k~nm}$ are described in the Appendix~\ref{A3}. Employing the evolution equation, the shear viscous correction to the distribution function can be expressed as, 
\begin{align}\label{2.11}
    \delta f_k\equiv\delta f_k^{\text{shear}}=\dfrac{\beta\,f_k^0{\tilde{f}}^0_k}{2\beta_{\pi}(u\!\cdot\!\Tilde{q}_k)}\Tilde{q}_k^{\alpha}\Tilde{q}_k^{\beta}\pi_{\alpha\beta},
\end{align}
We employ the shear viscous part of the distribution function as described in Eq.~(\ref{2.11}) in  Eqs.~(\ref{Ap})-(\ref{B1p}) to obtain the non-equilibrium corrections to the heavy quark drag and diffusion coefficients in the viscous medium. We consider longitudinal boost invariant expansion to model the hydrodynamical evolution of the QGP medium. The boost invariant expansion can be expressed within the Bjorken prescription~\cite{Bjorken:1982qr} by employing the Milne coordinates $(\tau,x,y,\eta_s)$ where $\tau=\sqrt{t^2-z^2}$ is the proper time and $\eta_s=\tanh^{-1}(z/t)$ is the space-time rapidity with $u^\mu=(1, 0, 0, 0)$ and $g^{\mu\nu}=(1,-1,-1,-1/\tau^2)$. Employing the Milne coordinates, the Eq.~(\ref{2.11}) gets simplified to,
\begin{align}\label{2.12}
    \delta f_k^{\text{shear}}=\frac{f_k^0{\tilde{f}}^0_k\,\mathfrak{s}}{\beta_\pi \omega_k T \tau} \left(\frac{\eta}{\mathfrak{s}}\right) \left[\frac{|{\bf q_k}|^2}{3}-(q_k)_z^2\right],
\end{align}
where we have used $\pi^{\mu\nu}\sigma_{\mu\nu}=4\eta/3\tau^2$ and $\eta$ is the shear viscosity of the QGP medium. The EQPM description of the entropy density $\mathfrak{s}$ of the QGP medium is described in Ref.~\cite{Bhadury:2019xdf}.
%
%%%%%%%%%%%%%%%%%%%%%%%%%%%%%%%%%%%%%%%
\section{Results and discussions}
%%%%%%%%%%%%%%%%%%%%%%%%%%%%%%%%%%%%%%%
%

%%%%%%%%%%%%%%%%%%%%%%% A vs p,T %%%%%%%%%%%%%%%%%%%%%%%%
\begin{figure*}
\includegraphics[scale=1]{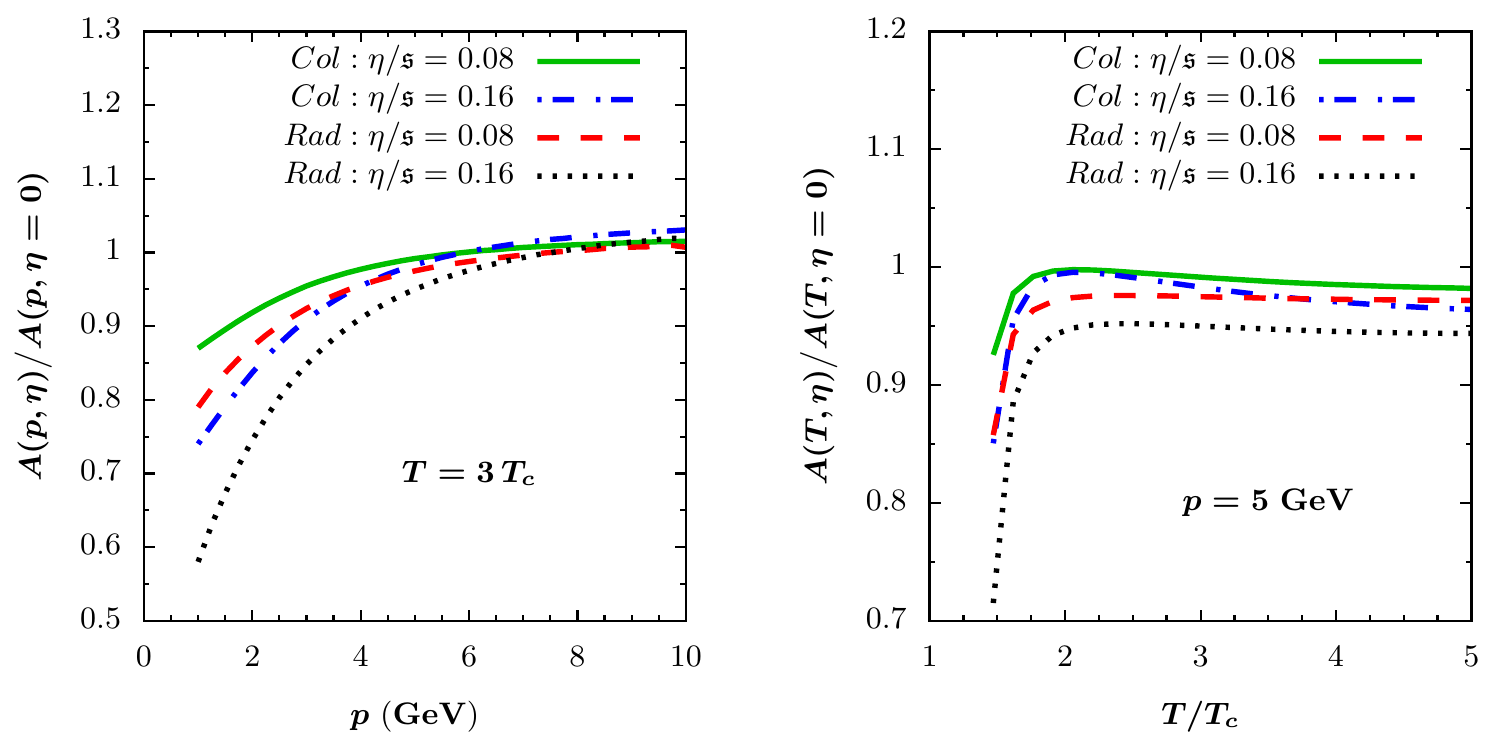} 
\caption{\small Drag coefficient $A(\eta)$ for the charm quark including shear viscous correction and scaled with its corresponding value for non-viscous case $A(\eta=0)$ as a function of its initial momentum (left panel) at $T=3\,T_c$ and as a function of QGP temperature (right panel) at $p=5$ GeV.}
\label{fig:3}
\end{figure*}
%%%%%%%%%%%%%%%%%%%%%%%%%%%%%%%%%%%%%%%%%%%%%%%%%%%%%%%%

%%%%%%%%%%%%%%%%%%%%%%% B0 vs p,T %%%%%%%%%%%%%%%%%%%%%%
\begin{figure*}
\includegraphics[scale=1]{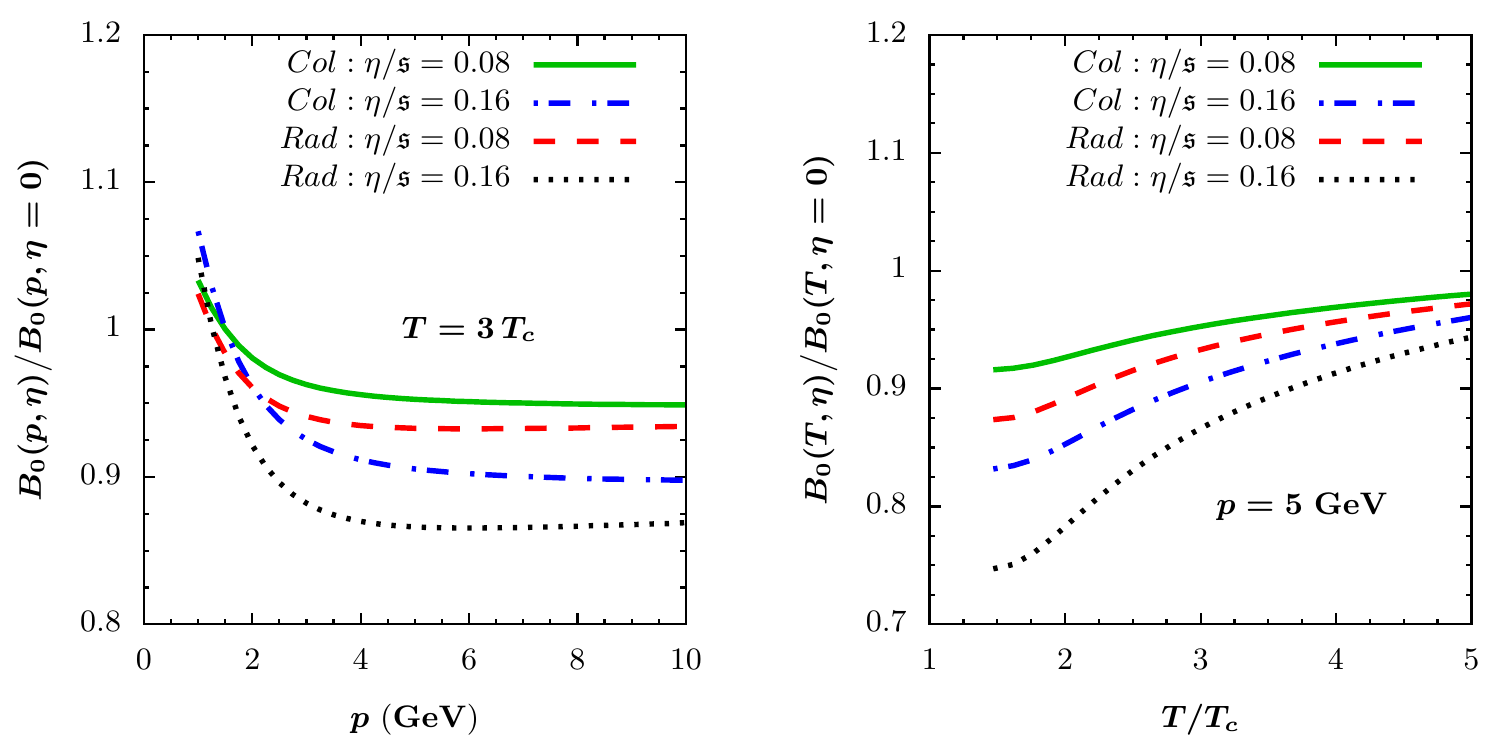} 
\caption{\small Transverse momentum diffusion coefficient $B_0(\eta)$ for the charm quark including shear viscous correction and scaled with its corresponding value for non-viscous case $B_0(\eta=0)$ as a function of its initial momentum (left panel) at $T=3\,T_c$ and as a function of temperature (right panel) at $p=5$ GeV.}
\label{fig:4}
\end{figure*}
%%%%%%%%%%%%%%%%%%%%%%%%%%%%%%%%%%%%%%%%%%%%%%%%%%%%%%%%

%%%%%%%%%%%%%%%%%%%%%%% B1 vs p,T %%%%%%%%%%%%%%%%%%%%%%%
\begin{figure*}
\includegraphics[scale=1]{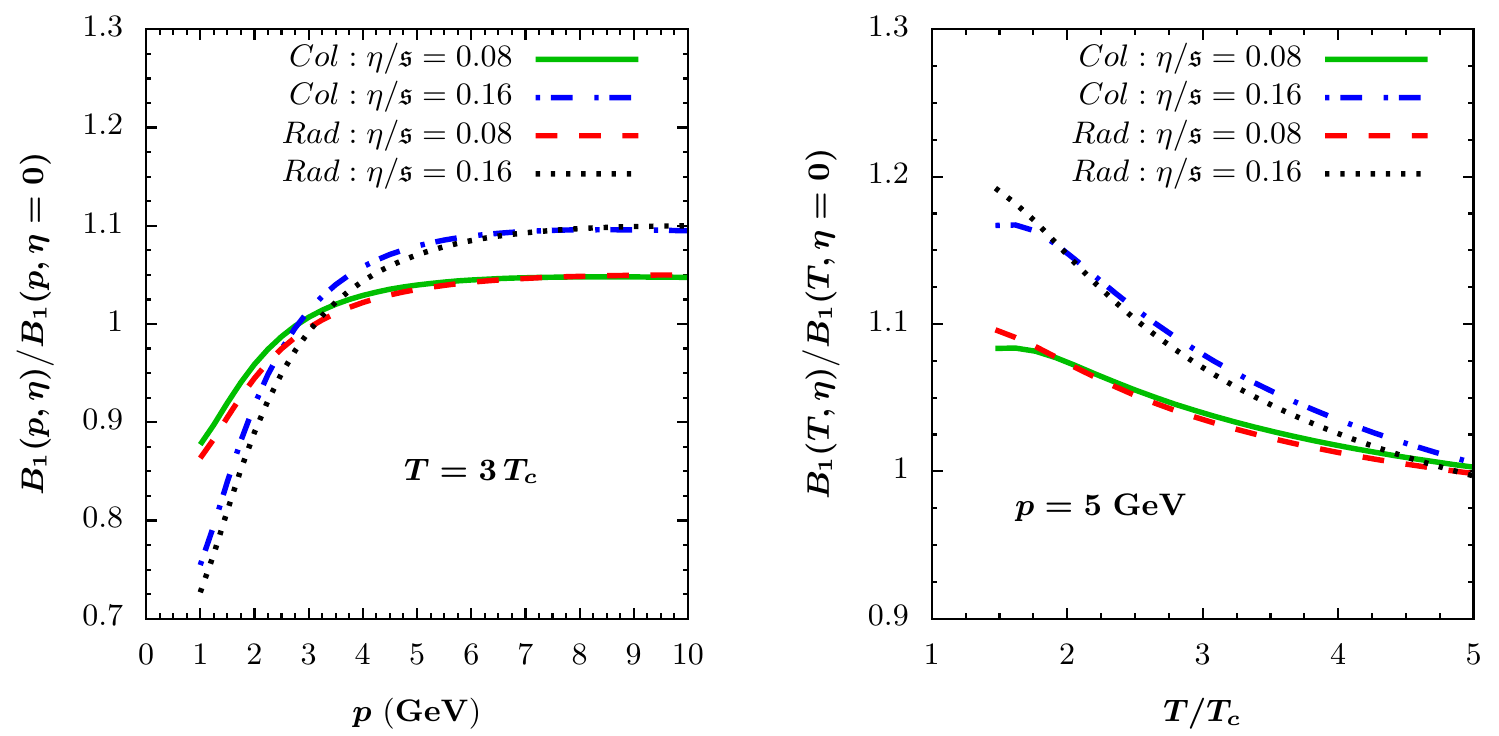} 
\caption{\small Longitudinal momentum diffusion coefficient $B_1(\eta)$ for the charm quark including shear viscous correction and scaled with its corresponding value for non-viscous case $B_1(\eta=0)$ as a function of its initial momentum (left panel) at $T=3\,T_c$ and as a function of temperature (right panel) at $p=5$ GeV.}
\label{fig:5}
\end{figure*}
%%%%%%%%%%%%%%%%%%%%%%%%%%%%%%%%%%%%%%%%%%%%%%%%%%%%%%%%

%%%%%%%%%%%%%%%%%%%%%%%%%%%%%%%%%%%%%%%%%%%%%%%%%%%%%%%%
\subsection{Heavy quark drag and momentum diffusion in the viscous medium}
%%%%%%%%%%%%%%%%%%%%%%%%%%%%%%%%%%%%%%%%%%%%%%%%%%%%%%%%
%

In the current analysis, we have studied the heavy quark transport coefficients within the viscous QGP considering medium-induced gluon emission of the charm quark in addition to the collisional process. 
For the quantitative analysis, we choose the mass of charm quark $m_c=1.3$ GeV, quark-hadron transition temperature $T_c = 170$ MeV for three flavors and the proper time $\tau = 0.25$ fm. We have studied the effect of viscous corrections for two values of the shear viscosity to entropy density ratio $\eta/\mathfrak{s}$ = $1/4\pi, \, 2/4\pi$. We have worked in the limit of massless light quarks with three flavors ($u,d,s$) and zero net baryon density ($\mu_{lq}=0$). Therefore, our study is valid in the regime where $m_{HQ}>>T>>m_{lq},\mu_{lq}$.

Fig.\,\ref{fig:3} (left panel) depicts the effects of shear viscous corrections on the momentum behaviour of the charm quark drag coefficient due to the collisional and radiative processes in the QGP medium. The drag coefficient of the heavy quark in the viscous QGP is critically dependent on its momentum along with the temperature of the medium. The momentum dependence of the drag coefficient due to the collisional and radiative processes can be described from Eqs.~(\ref{Ap}), ~(\ref{Fcm}), and~(\ref{rad}). It is observed that the shear viscosity reduces the heavy quark drag at low momentum regimes in contrast to the high momentum region. This could be realized from the interplay of two terms in Eq.~(\ref{Ap}) in the low and high momentum regimes while incorporating the viscous effects through Eq.~(\ref{2.12}).  The shear viscous correction is more prominent at low momenta of the charm quark around $p\approx1-3$ GeV, and an increase in the shear viscosity results in a decrease in the drag coefficient for both collisional and radiative processes. It is observed that the drag coefficient increases with an increase in $\eta/\mathfrak{s}$ at high momentum ($p\approx10$ GeV). 
Fig.\,\ref{fig:3} (right panel) displays the effect of the variation of the scaled drag coefficient $A(\eta)/A(\eta=0)$ as a function of the scaled QGP temperature $T/T_c$. 
It is seen that the shear viscous effect reduces the heavy quark drag coefficient throughout the relevant temperature range at $p=5$ GeV (Fig.\,\ref{fig:3}, right panel). This behaviour is mainly due to the negative contribution from the momentum factor $\left[\frac{|{\bf q_k}|^2}{3}-(q_k)_z^2\right]$ in $\delta f_k$ as described in Eq.~(\ref{2.12}). The shear viscous effect is more pronounced in the temperature regime near the transition temperature. This can be anticipated from the temperature behaviour of $\beta_\pi$ in the definition of $\delta f_k$ in Eq.~(\ref{2.12}). It is important to emphasize that the $\beta_{\pi}\propto T^4$ such that $\frac{\mathfrak{s}}{\beta_\pi T}\propto \frac{1}{T^2}$ in Eq.~(\ref{2.12}) (note that $\beta_\pi=\frac{4P}{5}$ for the ideal EoS, where $P$ is the pressure of the QGP medium).
Overall, it is important to emphasize that for charm quark momentum $p \approx 1$ GeV at $T=3\,T_c$ (left panel of Fig.\,\ref{fig:3}), maximum deviation to the ratio $A(\eta)/A(\eta=0)$ is observed where the drag coefficient ratio ranges from $\approx 0.85 - 0.75$ for collisional case and from $\approx 0.8 - 0.55$ for the radiative case with an increase in $\eta/\mathfrak{s}$ from $0.08$ to $0.16$. Qualitatively, similar behaviour is observed at low temperature $T=1.5\,T_c$ for the charm quark of momentum $p=5$ GeV (right panel of Fig.\,\ref{fig:3})

%%%%%%%%%%%%%%%%%%%%%%% EL vs p %%%%%%%%%%%%%%%%%%%%%%%%
\begin{figure*}
\includegraphics[scale=1]{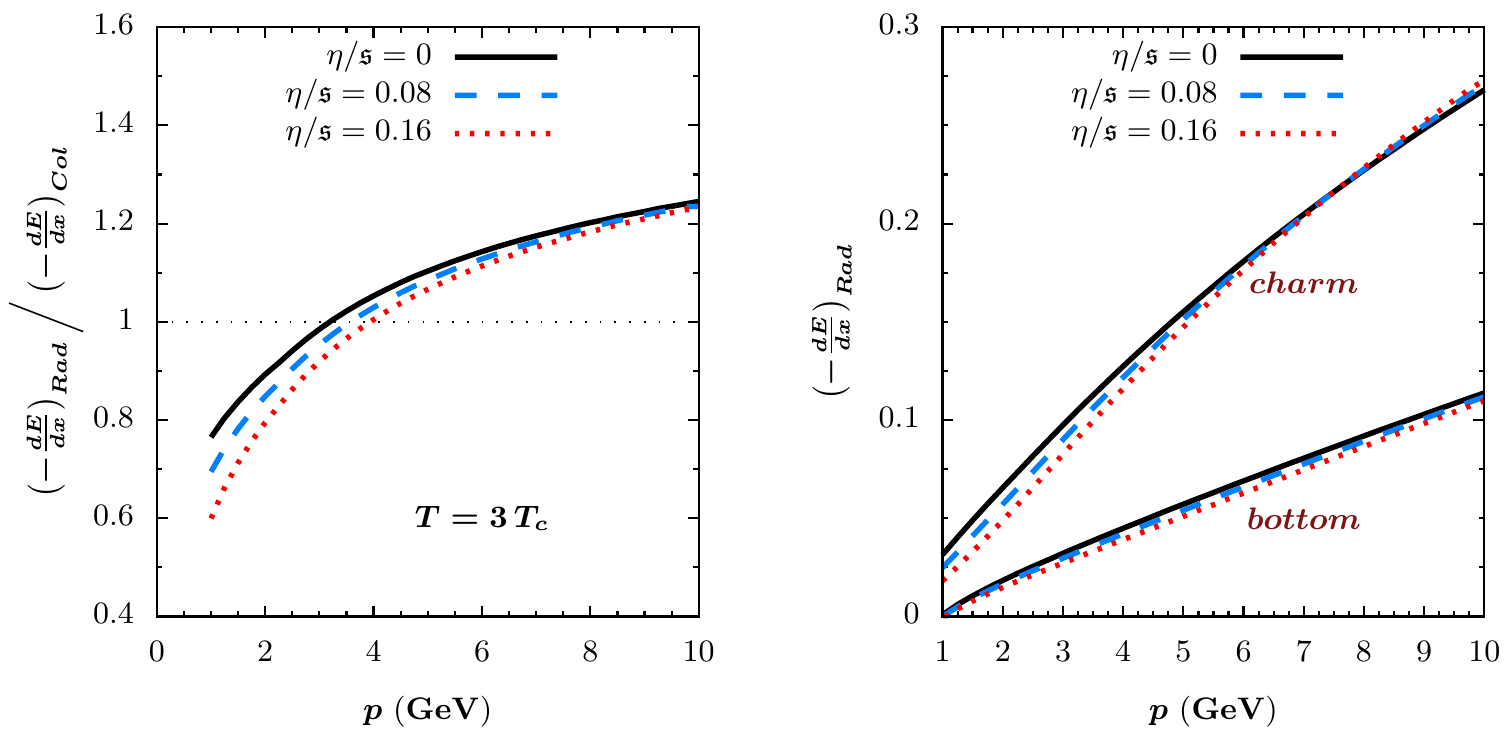} 
\caption{\small Ratio of the radiative to collisional differential energy loss for the charm quark at $3\,T_c$ (left panel). The differential radiative energy loss for the charm and the bottom quark at $3\,T_c$ (right panel).}
\label{fig:6}
\end{figure*}
%%%%%%%%%%%%%%%%%%%%%%%%%%%%%%%%%%%%%%%%%%%%%%%%%%%%%%%%

The momentum dependence of the transverse momentum diffusion coefficient ($B_0$) of the charm quark is depicted in Fig.\,\ref{fig:4} (left panel). In contrast to the drag coefficient, the transverse diffusion coefficient of the charm quark increases with the shear viscous correction at near $p\approx1$ GeV. For momenta $p\gtrsim3$ GeV, the viscous correction reduces the transverse diffusion coefficient. In Fig.\,\ref{fig:4}  (right panel), the transverse momentum diffusion coefficient is studied as a function of temperature. 
Both the momentum and temperature dependence indicate suppression of the transverse diffusion coefficient of the charm quark with an increase in shear viscosity. This suppression, however, is observed to be relatively more for the radiative process compared to the collision.

The longitudinal momentum diffusion coefficient ($B_1$) of the charm quark is shown in Fig.\,\ref{fig:5} (left panel). We observed that the viscous correction considerably reduces the coefficient at low momenta ($p\lesssim2$ GeV) affecting both collisional and radiative curves equally with the variation of $\eta/\mathfrak{s}$. For momenta $p\gtrsim6$ GeV, the longitudinal diffusion coefficient increases as compared to its value in the absence of shear viscosity, with higher $\eta/\mathfrak{s}$ resulting in a larger deviation.
The effect of $\eta/\mathfrak{s}$ on the charm quark longitudinal diffusion coefficient for $T=3\,T_c$ (left panel of Fig.\,\ref{fig:5}) is observed to be quite the opposite for low momenta ($p \approx 1$ GeV) in comparison to the high momenta ($p \approx 10$ GeV) for both collision and radiative cases.
The temperature dependence of the longitudinal momentum diffusion coefficient is depicted in Fig.\,\ref{fig:5}  (right panel). For the charm quark momentum of $p=5$ GeV in the temperature regime of $T<4\,T_c$, the ratio $B_1(\eta)/B_1(\eta=0)$ seems to increase with an increase in $\eta/\mathfrak{s}$ for both collision and radiative processes.

Following the same arguments for the temperature behaviour of heavy quark drag coefficient, the shear viscous effect (entering through the $\delta f_k$ with $\frac{\mathfrak{s}}{\beta_\pi T}\propto \frac{1}{T^2}$) to the diffusion coefficients is more visible in the low temperature regimes (right panels of Fig.\,\ref{fig:4} and Fig.\,\ref{fig:5}). The momentum dependence of $B_0$ and $B_1$ is described in Eq.~(\ref{B0p}) and Eq.~(\ref{B1p}), respectively. The viscous corrections will modify each terms in Eq.~(\ref{B0p}) and Eq.~(\ref{B1p}) such as $\langle\langle p'^{2} \rangle\rangle$, $\langle\langle ({\bf{p.p'}})^2\rangle\rangle$, $\langle\langle ({\bf{p.p'})}\rangle\rangle$, and $\langle\langle 1 \rangle\rangle$ via Eq.~(\ref{Fcm}) and Eq.~(\ref{rad}) by employing Eq.~(\ref{2.12}). This will affect the qualitative behaviour of the longitudinal and transverse diffusion coefficients in the viscous QGP medium.
%%%%%%%%%%%%%%%%%%%%%%%%%%%%%%%%%%%%%%%%%%%%%%%%%%%%%%%%
\subsection{Collisional and radiative energy loss in viscous medium}
%%%%%%%%%%%%%%%%%%%%%%%%%%%%%%%%%%%%%%%%%%%%%%%%%%%%%%%%
%

The differential energy loss is related to the drag coefficient of the heavy quark, which quantifies the resistance to the heavy quark motion due to the QGP constituents. The differential energy loss of the heavy quark can be expressed in terms of its drag coefficient as \cite{GolamMustafa:1997id},
\begin{equation}\label{EL}
    -\frac{dE}{dx} = p\,A(p).
\end{equation}
Fig.\,\ref{fig:6} (left panel) shows the ratio of the differential energy loss for the radiative (inelastic) process in comparison with the collisional (elastic) energy loss of the charm quark for different values of $\eta/\mathfrak{s}$ in the viscous QCD medium at temperature of $3\,T_c=510$ MeV. We observe that the elastic collision is the dominant mode of energy loss for the heavy quark in the low momenta regime (up to $p\approx3$ GeV), whereas beyond $p\approx 4$ GeV, radiative energy loss dominates. Increasing shear viscosity decreases the ratio of radiation to collisional energy loss for low momentum. However, the viscous corrections have a negligible effect at high momenta regimes ($p\gtrsim7$ GeV). 

The differential energy loss for the radiative process is shown in Fig.\,\ref{fig:6} (right panel) for the charm quark and the bottom quark at the $T=3\,T_c$. The suppression in the radiative energy loss of the bottom quark in comparison to the charm quark is due to the dead-cone effect, which prohibits the heavy quark from radiating gluon at a small angle. The higher the quark mass, the larger is the dead-cone angle, and less is the probability of the energy loss due to radiation. The shear viscous corrections are significant at low momentum ($p\approx 2$ GeV) for the charm quark radiative process. However, for the bottom quark, which is almost three times heavier than the charm quark, including non-equilibrium shear viscous corrections does not lead to a visible deviation from the equilibrium case when compared with the charm quark.
%
%%%%%%%%%%%%%%%%%%%%%%%%%%%%%%%%%%%%%%%%%%%%%%%%%%%%%
\section{Conclusion and Outlook}
%%%%%%%%%%%%%%%%%%%%%%%%%%%%%%%%%%%%%%%%%%%%%%%%%%%%%
%

In this article, we have investigated heavy quark transport and its energy loss by considering the collisional and radiative processes in the viscous QGP medium. The Brownian motion of the heavy quark in the hot QCD medium has been studied using the Fokker-Planck dynamics. The inelastic gluon radiation of the heavy quark, along with the elastic collision interactions with the medium constituents, have been included in the study of the transport coefficients, namely, drag and momentum diffusion coefficients. The thermal medium interactions are embedded in the analysis through EQPM effective degrees of freedom via temperature dependent effective fugacity parameters of quarks, antiquarks, and gluons.

We have estimated the viscous corrections to the momentum and temperature dependence of the charm quark drag and diffusion coefficients. The viscous corrections to the heavy quark transport coefficients enter through the quarks, antiquarks, and gluon momentum distribution functions. The shear viscous corrections to the distribution function employed in this analysis are obtained by solving the effective Boltzmann equation based on the EQPM framework. It is seen that the effects of viscous corrections on the drag and diffusion coefficients are larger for the radiative process in comparison with that to the collisional process in the expanding QGP medium, especially for the slow-moving charm quark and in the low temperature regimes. We have also estimated the charm quark collisional and radiative energy losses within the viscous QGP and studied their sensitivity to the shear viscosity. Further, we have observed suppression of the gluon radiation in the viscous QGP medium for the bottom quark in comparison to the charm quark due to the large mass of the bottom quark, whose thermalization time is comparatively large.

The viscous correction and realistic EoS effects to the heavy quark transport coefficients may affect the nuclear modification factor $R_{AA}$ and the collective flow coefficients in the heavy-ion collisions. We intend to investigate the phenomenological implications of these viscous corrections by modelling the expanding hot QCD medium with a $(3+1)$ dimensional relativistic hydrodynamic approach in the near future. The gluon radiation by the heavy quark in a magnetized medium will be another work to follow in the future.

\begin{acknowledgements}
A.S. thanks Himanshu Verma for help with python computation. M.K. would like to acknowledge the Indian Institute of Technology Gandhinagar for Institute postdoctoral fellowship. S.K.D acknowledges Jane Alam and Trambak Bhattacharyya for useful discussions. V.C. and S.K.D. acknowledge SERB Core Research Grant (CRG) [CRG/2020/002320].
\end{acknowledgements}

\appendix
%
%%%%%%%%%%%%%%%%%%%%%%%%%%%%%%%%%%%%%%%%%%%%%%%%%%%%%%%%
\section{Matrix element for heavy quark scattering}\label{A1}
%%%%%%%%%%%%%%%%%%%%%%%%%%%%%%%%%%%%%%%%%%%%%%%%%%%%%%%%
%
For the elastic $2\rightarrow 2$ collision (Fig.~\ref{Feyn_col}), the matrix element squared for the heavy quark interaction with light quark ($lq$), light antiquark ($l\bar{q}$) and gluon ($g$) take the following forms \cite{Combridge:1978kx, Shtabovenko:2020gxv, Shtabovenko:2016sxi, Mertig:1990an}:
\begin{widetext}
$(i)$ For the process  $HQ+lq/l{\bar q}\rightarrow HQ+lq/l{\bar q}$,
\begin{equation*}
   |{\mathcal{M}}_{(a)}|^2=\gamma_{HQ} \gamma_{lq/l{\bar q}} \left[\frac{64 \ \pi^2 \alpha^2}{9} \  \frac{(s-m_{HQ}^2)^2+(m_{HQ}^2-u)^2+2tm_{HQ}^2}{(t-m_D^2)^2}\right.
\end{equation*}
$(ii)$ For the process $HQ+g\rightarrow HQ+g$,
\begin{eqnarray*}
   |{\mathcal{M}}_{(b)}|^2&=&\gamma_{HQ} \gamma_{g} \left[\frac{64 \ \pi^2 \alpha^2}{9} \  \frac{(s-m_{HQ}^2)(m_{HQ}^2-u)+2m_{HQ}^2(s+m_{HQ}^2)}{(s-m_{HQ}^2)^2}\right],\\
   |{\mathcal{M}}_{(c)}|^2&=&\gamma_{HQ} \gamma_{g} \left[\frac{64 \ \pi^2 \alpha^2}{9} \  \frac{(s-m_{HQ}^2)(m_{HQ}^2-u)+2m_{HQ}^2(m_{HQ}^2+u)}{(m_{HQ}^2-u)^2}\right],\\
   |{\mathcal{M}}_{(d)}|^2&=&\gamma_{HQ} \gamma_{g} \left[32 \ \pi^2 \alpha^2 \  \frac{(s-m_{HQ}^2)(m_{HQ}^2-u)}{(t-m_{D}^2)^2}\right],\\
   \hspace{-2.25cm}{\mathcal{M}}_{(b)}{\mathcal{M}}_{(d)}^*&=&{\mathcal{M}}_{(b)}^*{\mathcal{M}_{(d)}}=\gamma_{HQ} \gamma_{g} \left[8 \ \pi^2 \alpha^2 \  \frac{(s-m_{HQ}^2)(m_{HQ}^2-u)+m_{HQ}^2(s-u)}{(t-m_{D}^2)(s-m_{HQ}^2)}\right],\\
   {\mathcal{M}}_{(c)}{\mathcal{M}}_{(d)}^*&=&{\mathcal{M}}_{(c)}^*{\mathcal{M}_{(d)}}=\gamma_{HQ} \gamma_{g} \left[8 \ \pi^2 \alpha^2 \  \frac{(s-m_{HQ}^2)(m_{HQ}^2-u)-m_{HQ}^2(s-u)}{(t-m_{D}^2)(m_{HQ}^2-u)}\right],\\
   {\mathcal{M}}_{(b)}{\mathcal{M}}_{(c)}^*&=&{\mathcal{M}}_{(b)}^*{\mathcal{M}_{(c)}}=\gamma_{HQ} \gamma_{g} \left[\frac{8 \ \pi^2 \alpha^2}{9} \  \frac{m_{HQ}^2(4m_{HQ}^2-t)}{(s-m_{HQ}^2)(m_{HQ}^2-u)}\right],
\end{eqnarray*}
\begin{equation*}
   |{\mathcal{M}}_{(2)}|^2=|{\mathcal{M}}_{(b)}|^2+|{\mathcal{M}}_{(c)}|^2+|{\mathcal{M}}_{(d)}|^2+2\mathcal{R}e\{{\mathcal{M}}_{(b)}{\mathcal{M}}_{(d)}^*\}+2\mathcal{R}e\{{\mathcal{M}}_{(c)}{\mathcal{M}}_{(d)}^*\}+2\mathcal{R}e\{{\mathcal{M}}_{(b)}{\mathcal{M}}_{(c)}^*\}.
\end{equation*}
\end{widetext}
%
%%%%%%%%%%%%%%%%%%%%%%%%%%%%%%%%%%%%%%%%%%%%%%%%%%%%%%
\section{Evaluation of the soft gluon 3-momentum integral for HQ radiative process}\label{A2}
%%%%%%%%%%%%%%%%%%%%%%%%%%%%%%%%%%%%%%%%%%%%%%%%%%%%%%%
%
The integral over 3-momentum of the radiated soft gluon excluding the kinematics and transport part is,
\begin{align}
      \mathcal{I} ({\bf k'})=& \int \frac{d^3 k'}{(2 \pi)^3 2 E_{k'}} \ \frac{12  g_s^2}{k'_\perp}\left(1+\frac{m_{HQ}^2}{s}e^{2y_{k'}}\right)^{-2} \nonumber\\
      &\times(1 + f(E_{k'})) \ \theta_1(E_p-E_{k'}) \ \theta_2(\tau-\tau_F),
\end{align}
where $k'\equiv(E_{k'},{\bf k'_\perp},k'_z)$. In terms of the rapidity of the radiated gluon, $y_{k'}=\frac{1}{2}\ln\left(\frac{E_{k'}+k'_z}{E_{k'}-k'_z}\right)$ we have,
\begin{align}
    \int_{-\infty}^{\infty} d^3 k' &= \int_{-\infty}^{\infty} d^2 k'_\perp \int_{-\infty}^{\infty} dk'_z\nonumber\\
    &= 2\pi \int_{0}^{\infty} k'_\perp \ dk'_\perp \int_{-\infty}^{\infty} E_{k'} \ dy_{k'}.
\end{align}
The theta function $\theta_2(\tau-\tau_F)$ demands $\tau>\tau_F$
where the the interaction time $\tau$ (inverse of interaction rate, $\Gamma=2.26 \ \alpha_s T$) is greater than the gluon formation time $\tau_F=\frac{\cosh \, y_k'}{k'_\perp}$ such that,
\begin{equation}\label{kperpmin}
     \Gamma^{-1}>\frac{\cosh \, y_{k'}}{k'_\perp} \implies k'_\perp > \Gamma \, \cosh \, y_{k'}.
\end{equation}
The theta function $\theta_1(E_p-E_{k'})$ restricts phase space for the heavy quark energy $E_p$ to be greater than the radiated gluon energy $E_{k'} = k'_\perp \cosh \, y_{k'}$ and we have,
\begin{equation}\label{kperpmax}
   \frac{E_p}{\cosh \, y_{k'}}>\frac{E_{k'}}{\cosh \, y_{k'}} \implies \frac{E_p}{\cosh \, y_{k'}} > k'_\perp.
\end{equation}
It is important to emphasize that the Eq.(\ref{kperpmin}) and (\ref{kperpmax}) sets the lower and upper bound respectively for the $k'_{\perp}$ integral. The Bose enhancement factor $(1 + f_g(E_{k'}))$ for the emitted soft gluon in the final state for the limiting case of $E_{k'} << T$ becomes,
\begin{equation}
    1 + f_g(E_{k'}) = 1+\frac{1}{e^{E_{k'}/T}-1} \approx \frac{T}{E_{k'}} = \frac{T}{k'_\perp \cosh \, y_{k'}}.
\end{equation}
So, the ${\bf k'}$ integral simplifies to, 
\begin{align}
    \mathcal{I} ({\bf k'}) = & \frac{3}{2 \pi^2} \ g_s^2 T \int_{\Gamma \, \cosh \, y_{k'}}^{E_p / \cosh \, y_{k'}} dk'_\perp \int_{-y}^{y} dy_{k'} \nonumber\\
    & \times \left(1+\frac{m_{HQ}^2}{s}e^{2y_{k'}}\right)^{-2} \frac{1}{k'_\perp \, \cosh \, y_{k'}},
\end{align}
where the limits of the rapidity integration is decided according to the pseudorapidity coverage of the detector.
%
%%%%%%%%%%%%%%%%%%%%%%%%%%%%%%%%%%%%%%%%%%%%%%%%%%%%%%%%%
\section{ Thermodynamic integrals}\label{A3}
%%%%%%%%%%%%%%%%%%%%%%%%%%%%%%%%%%%%%%%%%%%%%%%%%%%%%%%%%
%
The thermodynamic integrals $\Tilde{J}^{(r)}_{k~nm}$ and $\Tilde{L}^{(r)}_{k~nm}$ are respectively defined as follows,
\begin{align}
   \Tilde{J}^{(r)}_{k~nm}&=\frac{\gamma_k}{2\pi^2}\frac{(-1)^m}{(2m+1)!!}\int_{0}^\infty{d\mid{\Tilde{\bf p}}_k\mid}~\big(u\cdot\Tilde{p}_k \big)^{n-2m-r-1}\nonumber\\
&\times\big(\mid{\Tilde{\bf p}}_k\mid\big)^{2m+2} f^{0}_k\, \tilde{f}^0_k,\\
\Tilde{L}^{(r)}_{k~nm}&=\frac{\gamma_k}{2\pi^2}\frac{(-1)^m}{(2m+1)!!}\int_{0}^\infty{d\mid{\Tilde{\bf p}}_k\mid}~\frac{\big(u.\Tilde{p}_k\big)^{n-2m-r-1}}{\mid{\Tilde{\bf p}}_k\mid}\nonumber\\
    &\times\big(\mid{\Tilde{\bf p}}_k\mid\big)^{2m+2}f^0_k\tilde{f}^0_k. 
\end{align}
%\begin{widetext}
For the massless limit of the light quark, the thermodynamic integrals can be expressed in terms of the $PolyLog$ function as follows,
\begin{align}
    \Tilde{J}^{(1)}_{k~42} =&-\frac{2\,a_k\gamma_kT^5}{5\pi^2}\bigg[2\,\mathrm{PolyLog}~[4,-a_kz_k]\nonumber\\&
    -\frac{\delta\omega_k}{T}\,\mathrm{PolyLog}~[3,-a_kz_k]\bigg],
    \end{align}
\begin{align}
    \Tilde{L}^{(1)}_{k~42} &=-\frac{a_k\gamma_kT^4}{5\pi^2}~\mathrm{PolyLog}~[3,-a_kz_k].
\end{align}
%\end{widetext}

\bibliography{reference}{}

\end{document}